\documentclass[aip,apl,amsfonts,amssymb,amsmath,groupedaddress,singlecolumn]{revtex4}

\usepackage{graphicx}
\usepackage{amsmath}
\usepackage{amsfonts}
\usepackage{amssymb}
\usepackage{graphicx}
\usepackage{epstopdf}

\begin{document}
\title{Phonon transport across a Si-Ge interface: the role of inelastic bulk scattering}

\author{Jesse Maassen}
\email{jmaassen@dal.ca}
\affiliation{Department of Physics and Atmospheric Science, Dalhousie University, Halifax, Nova Scotia, Canada, B3H 4R2}
\author{Vahid Askarpour}
\affiliation{Department of Physics and Atmospheric Science, Dalhousie University, Halifax, Nova Scotia, Canada, B3H 4R2}

\begin{abstract}
Understanding phonon transport across heterojunctions is important to achieve a wide range of thermal transport properties. Using the McKelvey-Shockley flux method with first-principles modeling, we theoretically investigate the phonon transport properties of a Si-Ge interface with a focus on the role of inelastic bulk phonon processes. We observe significant inelastic scattering near the interface that redistributes the heat among the phonons as a result of non-equilibrium effects driven by the junction. These effects are most pronounced when the length of the junction is comparable to the average phonon mean-free-path. What controls these inelastic processes is elucidated.
\end{abstract}

\maketitle

\section{Introduction}
Developing material heterojunctions with a wide range of thermal transport properties is the focus of intense research, which can directly impact applications related to electronics, heat management and thermoelectrics \cite{Cahill2003,Cahill2014}. Although this topic has a long history, there still remain open fundamental questions regarding what microscopic processes control heat transport and dictate the thermal contact resistance, $R_C$.

Focusing on the case of interfaces comprised of semiconductors and/or insulators, in which phonons are the sole heat carriers, there have been many theoretical studies using a variety of methods to understand phonon transport and predict $R_C$. Simple models include the acoustic mismatch model (AMM) and diffuse mismatch model (DMM) \cite{Swartz1989}, where the AMM assumes wave coherence and specular scattering off a smooth interface, and the DMM assumes fully diffuse scattering off a rough interface. Thus, together both provide a range of $R_C$ values that can provide information about the quality of the interface when compared to experiment, although measurements do not always fall within this range.

Interfacial phonon transport has also been investigated using molecular dynamics (MD) \cite{Landry2009a,Landry2009b,Tomar2009,Puri2011,Chalopin2012,English2012,Merabia2012,Jones2013,Liang2014,Volz2014,Rashidi2015,Zhan2015,Gordiz2016,Gordiz2017,Feng2017,Zhou2017,Tao2017,Lee2018}, atomistic Green's function (AGF) \cite{Tian2012,Ong2015,Miao2016,Latour2017,Sadasivam2017} and the Boltzmann transport equation (BTE) \cite{Mansoor2011,Singh2011,Chen2013}. Each approach has its advantages and has been used to identify various phonon transport processes across heterojunctions. For example, MD captures all-order anharmonic phonon scattering, and was used to demonstrate the role of inelastic multi-phonon processes across the interface \cite{Volz2014,Zhan2015,Zhou2017}. AGF and MD naturally treat the interfacial phonon properties, which lead to non-bulk interface modes that have been proposed to mediate phonon transport \cite{Gordiz2016,Gordiz2017}. Detailed mode-resolved phonon transport properties \cite{Feng2017} and transmission coefficients \cite{Latour2017} have been obtained using AGF and MD. The BTE showed that inelastic bulk scattering can result in non-trivial temperature and heat current distributions near an interface \cite{Singh2011}. This may be a general feature of heterojunctions, but what exactly controls such inelastic processes and how this affects phonon transport has yet to be fully understood.

\begin{figure}	
\includegraphics[width=9.5cm]{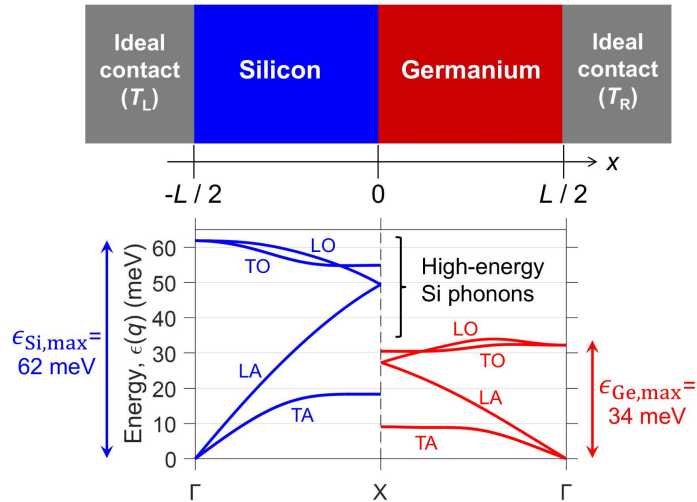}
\caption{Top: Si-Ge interface of length $L$, joined by two ideal contacts in thermodynamic equilibrium at different temperatures, $T_L$ and $T_R$. Bottom: Phonon dispersion of Si (left) and Ge (right).} \label{fig_model_structure}
\end{figure}

When there is a mismatch in the phonon energies, for example with Si and Ge as depicted in Fig. \ref{fig_model_structure}, inelastic phonon scattering is likely an important process. For example, since the maximum Si phonon energy ($\epsilon_{\rm Si,max}$) is nearly double that of Ge ($\epsilon_{\rm Ge,max}$), the high-energy Si phonons cannot conduct heat without inelastic scattering.

In this study, we investigate a Si-Ge interface focusing on the role and origin of inelastic bulk scattering in Si and Ge, using the McKelvey-Shockley flux method combined with density functional theory (DFT) and an ideal interface model. Our approach has the advantage of being computationally efficient and physically transparent, which helps provide new insights. We find the high-energy Si phonons do carry significant heat (more than in bulk) due to strong inelastic processes activated by non-equilibrium effects driven by the junction. While Si-Ge interfaces have been extensively investigated \cite{Tomar2009,Landry2009a,Landry2009b,Puri2011,Chalopin2012,Tian2012,Chen2013,Rashidi2015,Zhan2015,Gordiz2016,Zhou2017,Gordiz2017,Tao2017,Latour2017,Alkurdi2017,Larroque2018}, 
this work builds on prior studies \cite{Singh2011,Miao2016} to further elucidate the role of inelastic bulk scattering, a process that may be important in a broader class of heterojunctions.

\section{Simulated structure: Si-Ge interface} 
\label{sec:structure}
We consider a single Si-Ge interface of length $L$, with $L/2$ of both Si and Ge, as shown in Fig. \ref{fig_model_structure}. The interface is ideal, meaning atomically smooth with no strain or roughness present. The left side of Si and the right side of Ge are joined by two ideal contacts that are in thermodynamic equilibrium at different temperatures, $T_L$ and $T_R$. The contacts are assumed to be perfectly absorbing; any phonon in Si or Ge reaching a contact will not be reflected. In this study we take $T_L=301$ K and $T_R=300$ K, such that phonon transport is induced by a $\Delta T=1$ K.

\section{Theoretical Approach}
\label{sec:theory}
\subsection{McKelvey-Shockley flux method}
\label{sec:McK-S} 
The treatment of phonon transport inside the Si and Ge is carried out using the McKelvey-Shockley flux (McK-S) method. Originally developed for electron transport \cite{Mckelvey1961,Shockley1962}, it was subsequently adapted for phonon/thermal transport \cite{Maassen2015a,Maassen2015b}. The McK-S approach is a simple version of the BTE that captures ballistic and non-equilibrium effects, and can treat transport from the nano-scale to the macro-scale \cite{Maassen2015a,Maassen2015b,Maassen2016,Kaiser2017,Abarbanel2017}. Importantly, this method supports inelastic phonon scattering \cite{Abarbanel2017}.

The McK-S describes the spatial and temporal evolution of the forward- and reverse-moving phonon fluxes (along the transport direction, taken here as $x$). Instead of dealing with directed fluxes, it is possible to rewrite the governing equations directly in terms of temperature and heat current \cite{Abarbanel2017}. Temperature, in steady-state, is given by
\begin{align}
&D(\epsilon)\frac{\partial^2T(x,\epsilon)}{\partial x^2}=\frac{T(x,\epsilon)-T_0(x)}{\tau_Q(\epsilon)}, \label{temp_en} \\
&T_0(x)=\frac{\int_0^{\infty} T(x,\epsilon) \,C_V(\epsilon) \, /\, \tau_Q(\epsilon) \, {\rm d}\epsilon}{\int_0^{\infty} C_V(\epsilon) \, / \, \tau_Q(\epsilon) \, {\rm d}\epsilon}, \label{temp0} 
\end{align}
where $T(x,\epsilon)$ is the temperature resolved at each phonon energy, $T_0(x)$ the temperature of a fictitious local equilibrium distribution (originating from the use of the relaxation time approximation), $D(\epsilon)=\lambda(\epsilon)v_x^+(\epsilon)/2$ the $\epsilon$-resolved diffusion coefficient, $\tau_Q(\epsilon)=\lambda(\epsilon)/(2 v_x^+(\epsilon))$ the $\epsilon$-resolved heat relaxation time, $C_V(\epsilon)=\epsilon \, g(\epsilon)\, [\partial n_{\rm BE}/\partial T]_{T_{\rm ref}}$ the $\epsilon$-resolved heat capacity, $\lambda(\epsilon)$ the phonon mean-free-path for backscattering (average $x$-projected distance traveled before backscattering), $v_x^+(\epsilon)$ the average $x$-projected phonon velocity, $g(\epsilon)$ the phonon density of states, $n_{\rm BE}(\epsilon,T_{\rm ref})$ the Bose-Einstein distribution, and $T_{\rm ref}$ a reference temperature (300 K in this work). $\lambda(\epsilon)$ and $v_x^+(\epsilon)$ are defined in \cite{defs}. 

The expression for $T_0(x)$ is obtained by imposing conservation of energy (see Ref. \cite{Abarbanel2017}). In the form given by Eq. (\ref{temp0}), inelastic scattering is permitted (phonons can change energy during scattering). In some previous studies with the McK-S method \cite{Maassen2015a,Maassen2015b,Maassen2016,Kaiser2017}, only elastic scattering was permitted (phonons could not change energy during scattering); a discussion comparing elastic and inelastic scattering can be found in Ref. \cite{Abarbanel2017}. The treatment of scattering in this framework is at the level of the relaxation time approximation, which requires a scattering time to be specified, in this case 3-phonon anharmonic scattering calculated from first-principles.

Heat current, in steady-state, is expressed as
\begin{align}
I_Q(x,\epsilon) =-\kappa(\epsilon) \frac{\partial T(x,\epsilon)}{\partial x}, \label{iq_en}
\end{align}
where $\kappa(\epsilon)=C_V(\epsilon) D(\epsilon)$, is the $\epsilon$-resolved bulk thermal conductivity.

The boundary conditions (BCs) needed to solve for $T(x,\epsilon)$ at the left/right contacts are
\begin{align}
T(-L/2,\epsilon) &= T_L + \left. \frac{\lambda(\epsilon)}{2}\frac{{\rm d}T(x,\epsilon)}{{\rm d}x}\right|_{x=-L/2}, \label{left_bc} \\
T(L/2,\epsilon) &= T_R - \left. \frac{\lambda(\epsilon)}{2}\frac{{\rm d}T(x,\epsilon)}{{\rm d}x}\right|_{x=L/2}. \label{right_bc} 
\end{align}  
These mixed BCs are necessary to capture ballistic and non-equilibrium phonon effects \cite{Maassen2015a,Maassen2015b,Maassen2016,Kaiser2017}, and retrieve the traditional BCs (i.e. $T(-L/2,\epsilon)=T_L$ and $T(L/2,\epsilon)=T_R$) in the diffusive limit. Separate BCs are adopted for the Si-Ge junction (discussed below).

The effective, energy-integrated temperature and heat current are obtained using \cite{Abarbanel2017}
\begin{align}
T(x) = \frac{\int_0^{\infty} T(x,\epsilon) \,C_V(\epsilon)\,{\rm d}\epsilon}{\int_0^{\infty} C_V(\epsilon)\,{\rm d}\epsilon}, \label{temp} \\
I_Q(x)=\int_0^{\infty}I_Q(x,\epsilon)\,{\rm d}\epsilon. \label{iq}
\end{align}

This approach was shown to compare well to the more rigorous and computationally demanding BTE, particularly when solved using a full phonon dispersion \cite{Maassen2015a,Maassen2015b,Maassen2016,Kaiser2017,Abarbanel2017}.

\subsection{Ideal interface model}
To treat the Si-Ge interface, we adopt a simple interface model intended to capture the minimal thermal contact resistance between two dissimilar materials, assuming phonons transfer elastically across the junction. The motivation of this model, hereinafter referred to as the ideal interface model (IIM), is not to achieve better agreement with experiment but rather to (approximately) include the fundamental interfacial scattering that should exist in an ideal, smooth junction. A derivation, provided in Appendix \ref{app:interface_model}, is based on minimal assumptions: conservation of energy and detailed balance. AGF can provide a more accurate assessment of the fundamental transmission coefficient \cite{Latour2017}, but is more computationally expensive, particularly if both materials are not well lattice matched. Non-idealities (e.g. interfacial strain or defects) and other processes not considered here (e.g. inelastic interface scattering \cite{Hopkins2009}) will alter the transport properties. We refer interested readers to Appendix \ref{app:interface_model} for further details of the IIM and how it relates to the AMM and DMM.

With the IIM, the transmission coefficients are simply:
\begin{align}
\mathcal{T}_{i \rightarrow f}(\epsilon)  = \left\{ \begin{array}{cl} M_f(\epsilon) / M_i(\epsilon), & M_i(\epsilon) \geq M_f(\epsilon) \\
1, & M_i(\epsilon) < M_f(\epsilon) \end{array} \right. \label{trans_model}
\end{align} 
where $\mathcal{T}_{i\rightarrow f}(\epsilon)$ is the transmission probability of a phonon with energy $\epsilon$ traveling from material $i$ to $f$, and $M_{i/f}(\epsilon)$ is the distribution of modes (also referred to as the number of conducting channel per cross-sectional area) of the phonon states in material $i/f$. Phonons traveling from a material with few channels to a material with more channels ($M_f>M_i$) have a transmission of one, since there are enough final phonon states to accommodate all incoming phonons. However, the transmission is below one for the phonons impinging on a material with fewer number of channels ($M_f<M_i$), since there are not sufficient final states. See \cite{defs} for the definition of $M(\epsilon)$.

The transmission coefficients calculated from Eq. (\ref{trans_model}) are used in the following expression the for $\epsilon$-resolved thermal contact resistance (see Appendix \ref{app:interface_model}):
\begin{align}
R_C(\epsilon) &= \frac{1}{2} \left[  \left( \frac{1-\mathcal{T}_{1\rightarrow2}(\epsilon)}{\mathcal{T}_{1\rightarrow2}(\epsilon)} \right) R_1^{\rm ball}(\epsilon) +  \left( \frac{1-\mathcal{T}_{2\rightarrow1}(\epsilon)}{\mathcal{T}_{2\rightarrow1}(\epsilon)} \right) R_2^{\rm ball}(\epsilon) \right], \label{rc_en}
\end{align} 
where $R^{\rm ball}(\epsilon)=C_V(\epsilon) v_x^+(\epsilon)/2$ is the ballistic thermal resistance. Equation (\ref{rc_en}) is relatively general, assuming only elastic transport across the interface and detailed balance, and handles non-equilibrium phonon distributions. The IIM retrieves two known limits: $R_C=0$ in the case of two identical materials, and $R_C(\epsilon)=R^{\rm ball}/2$ when a material is joined with an ideal Landauer contact.

Using the IIM, given by Eqns. (\ref{trans_model})-(\ref{rc_en}), we obtain the following boundary conditions for the Si-Ge interface, assumed here to be at $x=l$:
\begin{align}
\Delta T(l,\epsilon)&=R_C(\epsilon) I_Q(l^{\pm},\epsilon), \label{bc1} \\
I_Q(l^+,\epsilon)&=I_Q(l^-,\epsilon), \label{rc2}
\end{align}
which states that the contact resistance will result in a discrete temperature drop across the interface $\Delta T(l,\epsilon)=T(l^-,\epsilon)-T(l^+,\epsilon)$ when a heat current flows, and that the heat current is continuous at the interface, respectively. The condition of continuous heat current across the interface, Eq. (\ref{rc2}), implies that there is no heat generation at the interface. Note that the effective thermal contact resistance is not obtained by an integration of $R_C(\epsilon)$ (or its inverse) over energy, but rather is extracted using $R_C=\Delta T(l)/I_Q(l)$ from the $\epsilon$-integrated, effective temperature and heat current.

\begin{figure}	
\includegraphics[width=11cm]{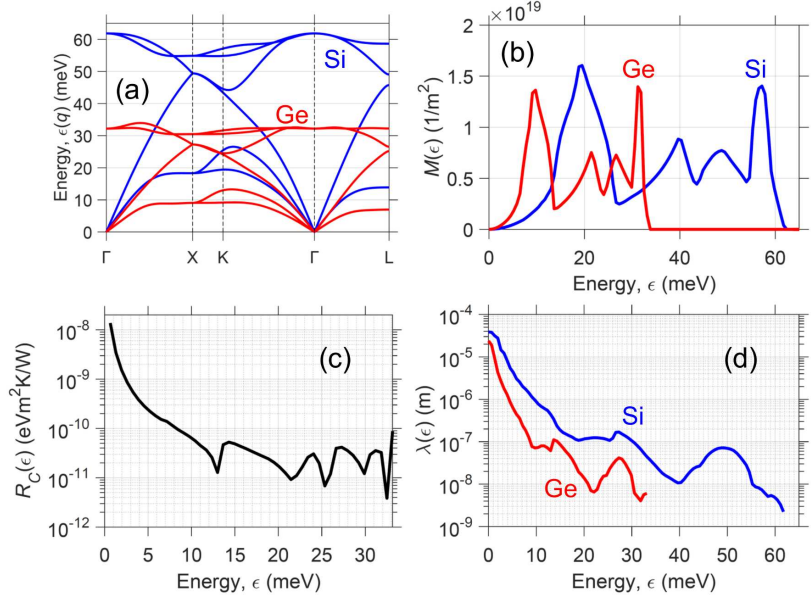}
\caption{(a) DFT-simulated phonon dispersion of Si and Ge. (b) Number of conducting channels per cross-sectional area, $M(\epsilon)$, of Si and Ge. (c) Energy-dependent thermal contact resistance, calculated using Eq. (\ref{rc_en}), for an ideal Si-Ge interface. (d) Mean-free-path for backscattering, $\lambda(\epsilon)$, of Si and Ge at 300 K, arising from anharmonic 3-phonon scattering obtained from DFT.} \label{fig_bulk_phonons}
\end{figure}

\section{Results} 
\label{sec:results}
\subsection{Bulk phonon properties of Si and Ge}
DFT was utilized to compute the phonon dispersions and 3-phonon anharmonic scattering rates for Si and Ge, which serve as input for our Si-Ge interface calculations (details of the DFT modeling are found in Appendix \ref{app:dft_details}). Fig. \ref{fig_bulk_phonons}a shows the calculated phonon dispersions, $\epsilon(q)$, for Si and Ge, which agree well with previous calculations \cite{SiGephonon}. The maximum phonon energies in Si and Ge are $\epsilon_{\rm Si,max}=62$ meV and $\epsilon_{\rm Ge,max}=34$ meV, respectively. From the $\epsilon(q)$, we compute the number of conducting channels per cross-section, $M_{\rm Si,Ge}(\epsilon)$, presented in Fig. \ref{fig_bulk_phonons}b. At low energies the $M_{\rm Si,Ge}(\epsilon)$ scale quadratically, as expected for linear dispersions. Fig. \ref{fig_bulk_phonons}c shows the energy-dependent thermal contact resistance calculated with Eqns. (\ref{rc_en})-(\ref{trans_model}). $R_C(\epsilon)$ is largest for the low-energy acoustic modes, and decreases sharply until $\sim$10 meV above which the resistance is relatively similar. The mean-free-path for backscattering, $\lambda(\epsilon)$, is extracted and presented in Fig. \ref{fig_bulk_phonons}d. $\lambda(\epsilon)$ is longest for the acoustic phonons, where 3-phonon scattering is less likely, reaching $\sim$30 $\mu$m before decaying rapidly to values below 100 nm for $\epsilon>$10 meV. A comparison of this DFT computed $\lambda(\epsilon)$ for Si to the previously reported $\lambda(\epsilon)$ in Ref. \cite{Maassen2015a}, based on phenomenological expressions for scattering, shows that both agree well and yield similar magnitude and energy dependence. The bulk thermal conductivities are calculated to be 160 W/m-K for Si and 51 W/m-K for Ge at 300 K, which compare well to previous simulations \cite{SiGekappa}.

\subsection{Si-Ge interface of fixed length (\textbf{\textit{L}}\,=\,200 nm)}
\begin{figure}	
\includegraphics[width=12.5cm]{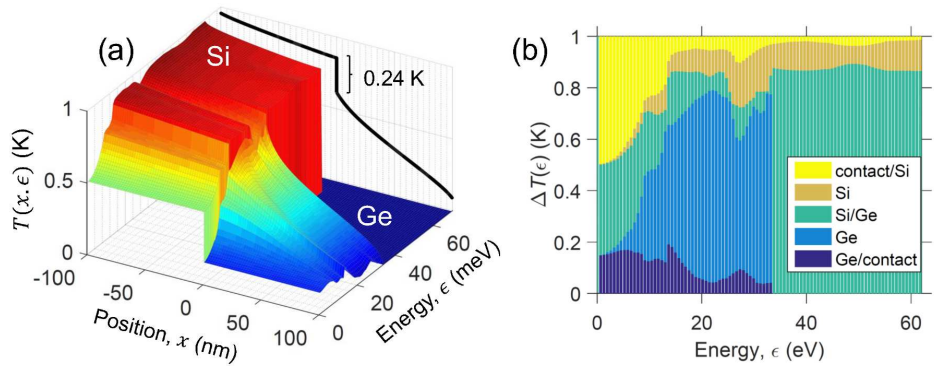}
\caption{(a) Temperature profile, $T(x,\epsilon)$, of a $L=200$ nm Si-Ge interface versus position, $x$, and phonon energy $\epsilon$. Effective temperature, $T(x)$, plotted as solid black line. (b) Temperature drop across the various sections of the structure. $T_L-T_R=1$ K.} \label{fig_temp}
\end{figure}

The case of $L=200$ nm is interesting since some high-energy phonons are diffusive ($\lambda(\epsilon)$$\ll$$L$), while simultaneously some other low-energy phonons are ballistic ($\lambda(\epsilon)$$\gg$$L$). Fig. \ref{fig_temp}a shows the temperature of the Si-Ge junction versus position, $x$, and phonon energy, $\epsilon$. Above 34 meV, temperature on the Ge side is not defined since there are no phonon states. Most of the temperature drop occurs on the Ge side compared to the Si side, since the former has a lower thermal conductivity. At the interface, there is a discrete temperature drop that varies with energy; it is largest for the low-energy acoustic phonons, where $R_C(\epsilon)$ is high (see Fig. \ref{fig_bulk_phonons}c). The low-energy acoustic phonons have a flat temperature profile, accompanied with a large temperature drop at the left/right contacts, which is a signature of ballistic phonon transport (as discussed in Refs. \cite{Maassen2015a,Maassen2015b,Maassen2016,Kaiser2017}). The high-energy Si phonons have a temperature at the $x=-L/2$ boundary approching $T_L$, indicating diffusive phonon scattering. Fig. \ref{fig_temp}b shows the temperature drop in various sections of the structure versus energy. At low $\epsilon$, $\Delta T$ is mostly localized at the interface/contacts (ballistic transport), while at higher $\epsilon$ much of the $\Delta T$ occurs across the Si and mostly the Ge (diffusive transport).

Using Eq. (\ref{temp}) we calculate the effective, energy-integrated temperature profile, plotted as a solid black line in Fig. \ref{fig_temp}a. We find a temperature drop of 0.24 K, which is relatively large compared to $T_L-T_R=1$ K. The total heat current is $x$-independent and equal to $I_Q=1.1\times10^8$ W/m$^2$, resulting in an effective $R_C=2.2\times10^{-9}$ Km$^2$/W with the IIM ($3.5\times10^{-9}$ Km$^2$/W with the DMM). These values are close to other reported calculations using the DMM ($4.6\times10^{-9}$ Km$^2$/W) \cite{Larroque2018}, the AMM ($\approx 5.5\times10^{-9}$ Km$^2$/W) \cite{Larroque2018} and MD ($\approx3.1\times10^{-9}$ Km$^2$/W) \cite{Landry2009b}. The IIM yields the lowest $R_C$. Surprisingly, the reported AMM value is larger than that of the DMM. A traditional solution of the 1D heat equation gives a linear temperature profile, however we find a non-linear $T(x)$, particularly across the Ge. As discussed later, this is a result of strong inelastic scattering near the interface.

\begin{figure}	
\includegraphics[width=12.5cm]{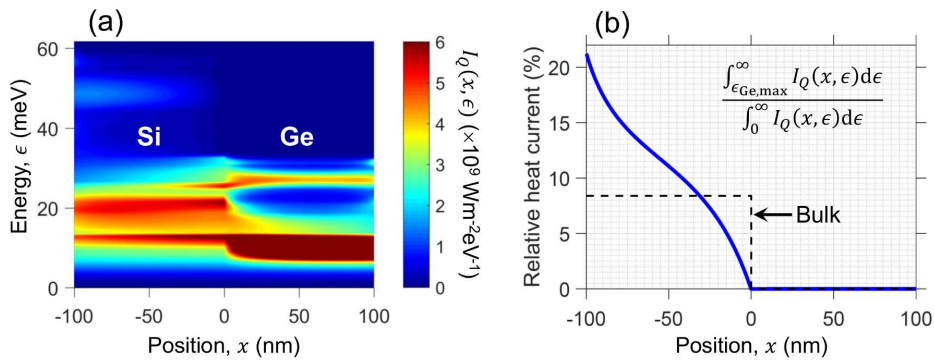}
\caption{(a) Heat current, $I_Q(x,\epsilon)$, of a $L=200$ nm Si-Ge interface. (b) Fraction of total heat current carried by phonons with energy larger than the maximum Ge phonon energy, $\epsilon_{\rm Ge,max}$. Black dashed line shows the  bulk value. $T_L-T_R=1$ K.} \label{fig_heat_current}
\end{figure}

Fig. \ref{fig_heat_current}a shows the heat current versus $x$ and $\epsilon$. $I_Q(x,\epsilon)$ flows at different energies in Si and Ge, which is only possible via inelastic scattering; elastic scattering would give an $x$-independent $I_Q(\epsilon)$. Similar features were observed in Ref. \cite{Singh2011}. At the interface, there are select energies, near 13 meV and 26 meV, that carry much of the heat from Si to Ge. These energies are found to have roughly equal number of conducting channels in both materials, $M_{\rm Si}(\epsilon)\approx M_{\rm Ge}(\epsilon)$, which results in small $R_C(\epsilon)$.

Focusing on the high-energy Si phonons, a non-zero $I_Q(x,\epsilon)$ is observed near 50 meV. Fig. \ref{fig_heat_current}b shows the fraction of the total heat current carried above the maximum Ge energy, which reaches over 20\% at the left boundary and decreases to zero at the interface (the fraction of heat current flowing above $\epsilon_{\rm Ge,max}$ in bulk Si is 8\%, shown as dashed line). Thus, a significant amount of heat is carried by the high-energy Si phonons, even though they are completely reflected at the interface. The fact that $I_Q(x,\epsilon)$ varies with $x$ indicates inelastic processes are rearranging the heat among the different phonons (i.e. the occupation of the phonon states), which we will illustrate next.

\begin{figure}	
\includegraphics[width=12.5cm]{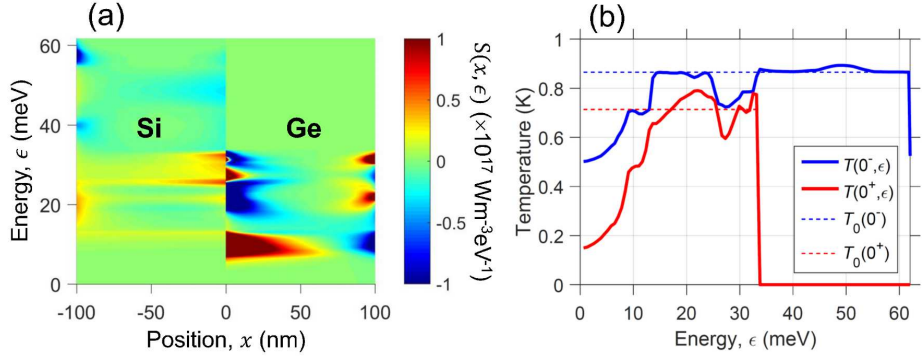}
\caption{(a) Heat sink/source, $S(x,\epsilon)$, of a $L=200$ nm Si-Ge junction. Positive/negative values correspond to heat added/removed from phonons. $\int_0^{\infty}S(x,\epsilon)\,{\rm d}\epsilon=0$. (b) Temperatures, $T(0,\epsilon)$ and $T_0(0)$, on both sides of the Si-Ge interface (Si/Ge side: $x=0^-/0^+$). $T_L-T_R=1$ K.} \label{fig_sinksource}
\end{figure}

With no external sink/source removing or adding heat to the structure, we obey the continuity equation for heat: $C_V\,\partial T(x)/\partial t = -\partial I_Q(x)/\partial x$. We can also write a similar equation that is phonon energy resolved: 
\begin{align}
C_V(\epsilon)\frac{\partial T(x,\epsilon)}{\partial t} &= -\frac{\partial I_Q(x,\epsilon)}{\partial x} + S(x,\epsilon), \label{continuity_eq}
\end{align}
in which we include a ``sink-source'' term, $S(x,\epsilon)$, describing how heat is being removed or added at each phonon energy, $\epsilon$. It can be shown that $\int_0^{\infty} S(x,\epsilon)\,{\rm d}\epsilon=0$, stating that heat is only redistributed among the phonons and that no net heat is removed or added. In steady-state, the left-hand side of Eq. (\ref{continuity_eq}) is zero, giving $S(x,\epsilon)=\partial I_Q(x,\epsilon)/\partial x$. Using Eqns. (\ref{temp_en}) and (\ref{iq_en}), we obtain:
\begin{align}
S(x,\epsilon)&=\frac{C_V(\epsilon)}{\tau_Q(\epsilon)}[T_0(x)-T(x,\epsilon)]. \label{sinksource}
\end{align} 
Fig. \ref{fig_sinksource}a presents $S(x,\epsilon)$ calculated using Eq. (\ref{sinksource}). The positive (red) and negative (blue) values correspond to heat being added and removed, respectively. In the vicinity of the interface, there is strong redistribution of heat, i.e. inelastic scattering. On the Si side, heat is removed from the high-energy phonons and tranferred to mid-energy phonons ($\sim$30 meV) that can travel into the Ge. Roughly speaking, as the high-energy phonons approach, they ``sense'' the interface and pump heat down to lower-energy phonons that can cross the junction. Since certain phonons transport most of the heat across the interface from Si to Ge, as shown in Fig. \ref{fig_heat_current}a, this creates a highly non-equilibrium distribution in the Ge side of the junction. As an attempt to drive the system towards equilibrium, in the Ge we observe strong inelastic scattering filling depleted states and emptying others. The redistribuion of heat near an interface occurs over a length scale comparable to $\lambda(\epsilon)$. We also see similar features near the left/right contacts. As previously discussed \cite{Maassen2015a,Maassen2015b}, the phonons injected from the contacts are in equilibrium with their originating thermal reservoir, but the phonons impinging on the contacts have a distribution controlled by the scattering that occurred inside the structure. The resulting phonon population near the contacts may be far from equilibrium, thus triggering inelastic processes.

To better understand what controls the heating/cooling pattern in Fig. \ref{fig_sinksource}a, in Fig. \ref{fig_sinksource}b we present the energy-resolved temperature, $T(0,\epsilon)$ (solid lines), and local equilibrium temperature, $T_0(0)$ (dashed lines), on both sides of the interface ($x=0^-$ for Si, and $x=0^+$ for Ge). According to Eq. (\ref{sinksource}), heat redistribution occurs whenever $T(x,\epsilon)\neq T_0(x)$; heat is added when $T_0(x)-T(x,\epsilon)>0$, and removed when $T(x,\epsilon)-T_0(x)>0$. With this insight, Fig. \ref{fig_sinksource}b shows which phonon energies will be heated/cooled, which correlates with the pattern in Fig. \ref{fig_sinksource}a. The low-energy acoustic modes are far from the equilibrium temperature, because they are near ballistic and do not efficiently scatter, however their value of $C_V(\epsilon)/\tau_Q(\epsilon)$ is small. Note that under near-equilibrium conditions $T(x,\epsilon)$ is independent of $\epsilon$, thus an energy dependence is indicative of a non-equilibrium phonon distribution. In summary, interfaces drive the phonon population out of equilibrium, which in turn activate inelastic scattering.

\subsection{Si-Ge interface with varying \textbf{\textit{L}}}
\begin{figure}	
\includegraphics[width=14.5cm]{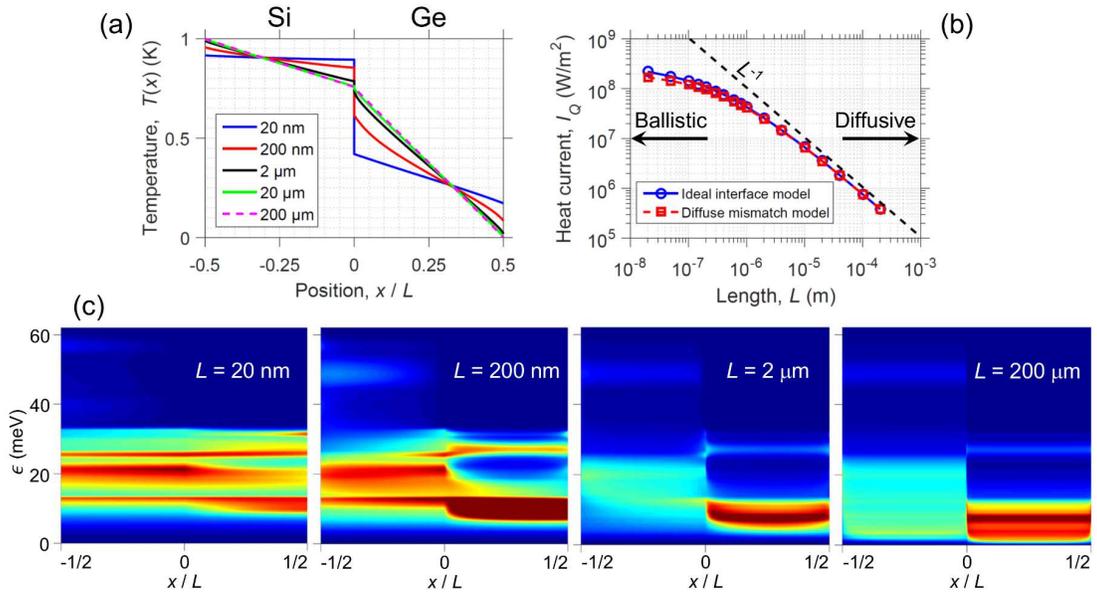}
\caption{Temperature profile, $T(x)$, and heat current, $I_Q$, of Si-Ge interfaces as a function of length $L$. (c) Energy-resolved heat current, $I_Q(x,\epsilon)$ for $L=20$ nm, 200 nm, 2 $\mu$m and 200 $\mu$m. $T_L-T_R=1$ K.} \label{fig6}
\end{figure}

Fig. \ref{fig6}a shows the temperature profile across the Si-Ge junction with $L$ ranging from 20 nm to 200 $\mu$m. With $L=20$ nm most of the temperature drop occurs at the left/right contacts and the inteface, indicating near-ballistic transport. With $L=200$ $\mu$m we retrieve the traditional heat equation solution, a linear $T(x)$ with vanishing temperature drops at the interface and contacts. Interestingly a distinct non-linearity is observed with $L=200$ nm, since this is the length scale over which inelastic scattering is strongest. With smaller $L$ scattering is sparse leading to a linear and flat $T(x)$ when approaching the ballistic limit, as discussed in \cite{Maassen2015a}. With larger $L$ most of the scattering occurs close to the interface/contacts and the phonons inside each material are near equilibrium resulting in a linear $T(x)$, as predicted by the classical heat equation. The non-linearity in $T(x)$, observed at intermediate length, arises from inelastic scattering, since assuming only elastic scattering (using the same approach) results in a linear $T(x)$ \cite{Maassen2015a}.  

The heat current versus $L$ is plotted in Fig. \ref{fig6}b, comparing the IIM to the DMM. For $L>20$ $\mu$m the heat current follows a $L^{-1}$ trend, as expected in the diffusive limit. Below $L=1$ $\mu$m the heat current rolls off, falling below the traditional scaling, as phonon transport becomes quasi-ballistic and approaches the ballistic limit. Fig. \ref{fig6}c presents the energy-resolved heat current for various $L$. Note how each plot is visually different until reaching $\sim$20 $\mu$m, above which $I_Q(x,\epsilon)$ for a given $x$ is the same as bulk. With $L=20$ nm, the heat current flows at roughly the same energies in both Si and Ge, since inelastic scattering occurs over a longer length scale (as discussed above). With $L=200$ nm, we observe the transition of heat current carried by certain energies in Si to others in Ge. For longer $L$, the energy-dependence of the heat currrent is mostly uniform in $x$ (away from the interface and contacts).

\section{Conclusions}
\label{sec:conclusions}
We theoretically analyzed phonon transport across a Si-Ge interface using the McKelvey-Shockley flux method combined with DFT materials modeling, which captures ballistic, non-equilibrium and inelastic scattering effects. This study focused mainly on the role of inelastic bulk phonon scattering that occurs near a heterojunction. We started by investigating a 200 nm long Si-Ge junction, in which we simultaneously observed diffusive transport of high-energy phonons and ballistic transport of low-energy phonons. Our analysis showed that strong inelastic scattering exists near the interface and the contacts. This originates from the non-equilibrium phonon distribution that is induced by the presence of the interface; heat is conducted across the junction at select phonon energies while other phonons are completely reflected (e.g. the high-energy Si phonons), which drive the system away from equilibrium. 

As a result of these inelastic processes, there is significant redistribution of heat among the phonons near an interface. This enables the high-energy Si phonons to carry significant heat, which are ``cooled'' within roughly one mean-free-path of the junction by pumping heat down to mid-energy Si phonons that can transfer into the Ge. What activates and controls these processes was elucidated. Moreover, a non-linear temperature profile is observed in the regions where inelastic scattering is significant. We also demonstrated how the phonon transport properties vary with the length of the junction, $L$, from 20 nm to 200 $\mu$m. We observe the ballistic to diffusive transition and show how this impacts the temperature and heat current distributions.

For our treatment of the Si-Ge interface we adopted a simple interface model intended to (approximately) capture the minimum thermal resistance of an ideal smooth junction. The physics of this model, and its connection to others including the AMM and DMM, were discussed, and an expression for thermal contact resistance was provided. 

Lastly, while our treatment is not comprehensive, for example not including interfacial phonon modes or inelastic phonon transport across the interface, it does highlight the importance of inelastic bulk scattering and non-equilibrium physics, which are processes that likely play a role in many other heterointerfaces beyond Si-Ge.

\acknowledgements
This work was partially supported by DARPA MATRIX (Award No. HR0011-15-2-0037) and NSERC (Discovery Grant RGPIN-2016-04881).

\appendix
\section{Thermal contact resistance and the ideal interface model}
\label{app:interface_model}
Consider a heterojunction at $x=l$, with the material to the left ($x\leq l^-$) labeled as $1$ and the material to the right ($x\geq l^+$) labeled as $2$. Assuming phonons travel elastically across the interface, a scattering matrix relating the incoming heat fluxes to the outgoing heat fluxes can be constructed:
\begin{align}
\begin{bmatrix} 
I_Q^+(l^+) \\ I_Q^-(l^-) 
\end{bmatrix}
=
\begin{bmatrix} 
\mathcal{T}_{1\rightarrow2} & 1-\mathcal{T}_{2\rightarrow1} \\
1-\mathcal{T}_{1\rightarrow2} & \mathcal{T}_{2\rightarrow1} 
\end{bmatrix}
\begin{bmatrix} 
I_Q^+(l^-) \\ I_Q^-(l^+) 
\end{bmatrix}, \label{scat_mat1}
\end{align}
where $I_Q^{\pm}(x,\epsilon)$ are the directed heat fluxes (calculated with the McK-S method in this study), and $\mathcal{T}_{i \rightarrow f}(\epsilon)$ is the transmission probability of a phonon traveling from the initial material $i$ to the final material $f$. Note that in this section the explicit $\epsilon$ dependence of quantities will often be dropped for clarity (but is implicitly assumed). Using this scattering matrix we first derive an expression for thermal contact resistance, $R_C$, then discuss the particular choice in transmission coefficients that make up the ideal interface model (IIM).

We begin by relating the directed heat fluxes to the net heat current and temperature, and deriving several useful properties. The directed heat fluxes are written as $I_Q^{\pm}(x)=I_{Q,{\rm ref}}^{\pm}+\delta I_Q^{\pm}(x)$, where $I_{Q,{\rm ref}}^+=I_{Q,{\rm ref}}^-$ are the heat fluxes of an equilibrium phonon distribution at a constant reference temperature $T_{\rm ref}$, and $\delta I_Q^{\pm}(x)$ are corrections to the background fluxes. The heat current is $I_Q(x)=I_Q^+(x)-I_Q^-(x)$, and temperature is $T(x)=(T^+(x)+T^-(x))/2$, where $T^{\pm}(x)=[I_Q^{\pm}(x)-I_{Q,{\rm ref}}^{\pm}(x)]/K^{\rm ball}+T_{\rm ref}$ is the temperature of the forward/reverse phonon streams, $K^{\rm ball}(\epsilon)=C_V v_x^+/2=\epsilon M (\partial n_{\rm BE}/\partial T)|_{T_{\rm ref}} / h$  is the ballistic thermal conductance, $C_V(\epsilon)$ is the heat capacity, $v_x^+(\epsilon)$ is the average $x$-projected velocity ($x$ being the transport direction), $M(\epsilon)$ is the distribution of modes, and $n_{\rm BE}(\epsilon,T)$ is the Bose-Einstein distribution (see Ref. \cite{Maassen2015a} for more details on these quantities). We assume small temperature variations, such that $|T(x)-T_{\rm ref}|\ll T_{\rm ref}$.

Next, a condition that no net heat current flows in equilibrium is imposed on Eq. (\ref{scat_mat1}). Using Eq. (\ref{scat_mat1}), we find
\begin{align}
&I_Q^+(l^+) - I_Q^-(l^+) \nonumber \\ &= \mathcal{T}_{1\rightarrow2} I_Q^+(l^-) - \mathcal{T}_{2\rightarrow1} I_Q^-(l^+) \nonumber \\
&= \mathcal{T}_{1\rightarrow2} I_{Q,{\rm ref}}^+(l^-) - \mathcal{T}_{2\rightarrow1} I_{Q,{\rm ref}}^-(l^+) \nonumber \\
&= 0. \label{heat_current1}
\end{align}
where $I_Q^{\pm} \rightarrow I_{Q,{\rm ref}}^{\pm}$ in equilibrium. Using $I_{Q,{\rm ref}}^{\pm}=\epsilon M n_{\rm BE}/h$, leads to the following condition relating $\mathcal{T}_{1\rightarrow2}$ and $\mathcal{T}_{2\rightarrow1}$:
\begin{align}
M_1 \, \mathcal{T}_{1\rightarrow2} = M_2 \, \mathcal{T}_{2\rightarrow1}, \label{detailed_bal2}
\end{align}
or equivalently,
\begin{align}
K_1^{\rm ball} \, \mathcal{T}_{1\rightarrow2} = K_2^{\rm ball} \, \mathcal{T}_{2\rightarrow1}. \label{detailed_bal1}
\end{align}
Note that this same condition is obtained if we had started with $I_Q(l^-)=0$ instead of $I_Q(l^+)=0$. Equations (\ref{detailed_bal2})-(\ref{detailed_bal1}) indicate that only $\mathcal{T}_{1\rightarrow2}$ must be specified, and that $\mathcal{T}_{2\rightarrow1}$ is automatically determined by detailed balance (or vice versa).

By adding both equations defined by \ref{scat_mat1}, one finds $I_Q(l^-)=I_Q(l^+)$, stating that the heat current is continuous across the interface. By subtracting $I_Q^-(l^-)$ from the first equation in \ref{scat_mat1}, the following equivalent expressions for heat current are found:
\begin{align} 
I_Q(l^{\pm})&=\mathcal{T}_{1\rightarrow2} I_Q^+(l^-) - \mathcal{T}_{2\rightarrow1} I_Q^-(l^+) \label{heat_current2} \\
&=\mathcal{T}_{1\rightarrow2} K_1^{\rm ball}T^+(l^-) - \mathcal{T}_{2\rightarrow1} K_2^{\rm ball} T^-(l^+) \label{heat_current3} \\
&=\mathcal{T}_{1\rightarrow2} K_1^{\rm ball}\left[ T^+(l^-) - T^-(l^+) \right] \label{heat_current4} \\
&=\mathcal{T}_{2\rightarrow1} K_2^{\rm ball}\left[ T^+(l^-) - T^-(l^+) \right], \label{heat_current5}
\end{align}
where Eq. (\ref{detailed_bal1}) was used.

We can now derive an expression for thermal contact resistance, $R_C$. By adding $I_Q^-(l^+)$ and $I_Q^+(l^-)$ to both sides of the first and second equations in \ref{scat_mat1}, respectively, and manipulating we arrive at expressions for temperature on both sides of the junction:
\begin{align}
T(l^+) &= T^-(l^+) + I_Q(l^{\pm})/(2 K_2^{\rm ball}), \label{temp_lplus} \\
T(l^-) &= T^+(l^-) - I_Q(l^{\pm})/(2 K_1^{\rm ball}). \label{temp_lminus} 
\end{align}
The discrete temperature drop across the interface is $\Delta T=T(l^-)-T(l^+)$. Subtracting Eq. (\ref{temp_lplus}) from Eq. (\ref{temp_lminus}), and using Eqns. (\ref{heat_current4})-(\ref{heat_current5}), we find:
\begin{align}
\Delta T &= \left[ \frac{(1-\mathcal{T}_{1\rightarrow2}) + (1-\mathcal{T}_{2\rightarrow1})}{\mathcal{T}_{1\rightarrow2} K_1^{\rm ball} + \mathcal{T}_{2\rightarrow1} K_2^{\rm ball}} \right] I_Q(l^{\pm}). \label{alter_rc} 
\end{align} 
Defining the thermal contact resistance as $R_C=\Delta T/I_Q(l^{\pm})$, we obtain our final expression:
\begin{align}
R_C &= \frac{1}{2}\left[ \frac{(1-\mathcal{T}_{1\rightarrow2})}{\mathcal{T}_{1\rightarrow2}} R_1^{\rm ball} + \frac{(1-\mathcal{T}_{2\rightarrow1})}{\mathcal{T}_{2\rightarrow1}} R_2^{\rm ball} \right] \label{rc} 
\end{align} 
where $R^{\rm ball}(\epsilon) = 1/K^{\rm ball}(\epsilon)$ is the ballistic thermal resistance. Note that Eq. (\ref{rc}) is rather general; its derivation is based on the assumption of elastic transfer of phonons across the interface and the principle of detailed balance. No assumption on the phonon dispersion was made, and this expression can be easily evaluated with a first-principles full phonon dispersion. Lastly, Eq. (\ref{rc}) is applicable under non-equilibrium phonon conditions.

Next we specify the particular choice in $\mathcal{T}_{i \rightarrow f}$ that will constitute the IIM. The purpose of this model is to (approximately) capture the minimal thermal contact resistance between two dissimilar materials under the assumption of an ideal, smooth interface. For this purpose, we ask ourselves what is the largest possible interfacial transmission coefficient? A transmission of one is allowed, but only for phonons flowing along one direction. Assuming that $M_1<M_2$, then $\mathcal{T}_{1 \rightarrow 2}=1$ is permitted, but according to Eq. (\ref{detailed_bal2}), $\mathcal{T}_{2 \rightarrow 1}=M_1/M_2<1$ by detailed balance. $\mathcal{T}_{2 \rightarrow 1}=1$ is not allowed, since this would result in $\mathcal{T}_{1 \rightarrow 2}=M_2/M_1>1$. If we now assume that $M_1>M_2$, then $\mathcal{T}_{2 \rightarrow 1}=1$ and $\mathcal{T}_{1 \rightarrow 2}=M_2/M_1<1$. When $M_1=M_2$, both transmissions are one, $\mathcal{T}_{1 \rightarrow 2}=\mathcal{T}_{2 \rightarrow 1}=1$.

\begin{figure}	
\includegraphics[width=9cm]{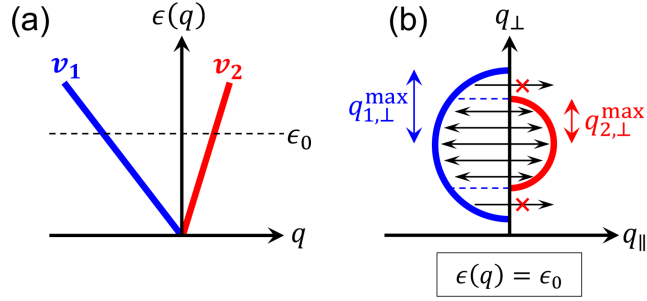}
\caption{(a) Linear phonon dispersions of two materials, $\epsilon(q)=\hbar v_{1,2} q$, where $v_1<v_2$. (b) Iso-energy surface in reciprocal space for both materials when $\epsilon(q)=\epsilon_0$. $q_{\perp}$ and $q_{\parallel}$ are the reciprocal lattice vectors perpendicular and parallel to the interface plane, respectively.} \label{fig7}
\end{figure}

This model retrives two known limits: {\it 1) Two identical materials.} In the case of two identical materials, the IIM states that $\mathcal{T}_{1 \rightarrow 2}=\mathcal{T}_{2 \rightarrow 1}=1$, and Eq. (\ref{rc}) gives $R_C=0$, as one would physically expect. {\it 2) Material interfaced with a Landauer contact.} An ideal thermal reservoir, sometimes referred to as a Landauer contact, is assumed to have a very large number of channels compared to the material. Labeling the material as $1$ and the Landauer contact as $2$, a careful evaluation of Eq. (\ref{rc}) yields $R_C\rightarrow R_1^{\rm ball}/2$ when $M_1\ll M_2$, a well-known result from the Landauer approach for phonon \cite{Maassen2015a} and electron transport \cite{Datta_book}.

As an example, let us consider an interface formed by two materials with linear phonon dispersions, $E(q)=\hbar v q$, where $v$ is the group velocity (here we assume $v_1<v_2$) and $q$ the phonon wavenumber (see Fig. \ref{fig7}a). The distribution of modes for a linear dispersion is $M(\epsilon)=3\pi \epsilon^2/(h^2 v^2)$ (factor of 3 is for three polarizations) \cite{Jeong2011}. According to the IIM, $\mathcal{T}_{1 \rightarrow 2}=v_1^2/v_2^2$ and $\mathcal{T}_{2 \rightarrow 1}=1$, which gives $R_C = (R_2^{\rm ball}-R_1^{\rm ball})/2$. The reason why the transmission from 1 to 2 is below one is exemplified in Fig. \ref{fig7}b, showing the iso-energy surface of the phonon dispersion in $q$-space for a specific energy $\epsilon_0$. With an ideal interface, transverse momentum / wavenumber ($q_{\perp}$) is conserved; such transitions are depicted by horizontal arrows in Fig. \ref{fig7}b. The maximum possible transverse wavenumber in 1 is larger than in 2, $q_{1,\perp}^{\rm max}>q_{2,\perp}^{\rm max}$. The incoming phonons from 1 can transfer into 2 if $q_{1,\perp}\leq q_{2,\perp}^{\rm max}$, but are reflected if $q_{1,\perp}> q_{2,\perp}^{\rm max}$ since there are no available final states in 2. Oppositely, all incoming phonons from 2 can transfer into 1, since $q_{2,\perp}< q_{1,\perp}^{\rm max}$ always holds.

The phonon momentum vector is altered when traveling from one material to another. In the linear dispersion example, phonon propagation occurs normal to any point on the iso-energy surface of the $\epsilon(q)$ shown in Fig. \ref{fig7}b. An incoming phonon with $q_{1,\perp} = q_{2,\perp}^{\rm max}$, will approach the interface with a moderate angle, but after crossing into material 2 will exit traveling parallel to the interface. As mentioned above, any incoming phonon with $q_{1,\perp} > q_{2,\perp}^{\rm max}$ will be backscattered. Hence, we can define a critical angle above which phonons are reflected, and below which phonons are refracted. Thus, this model is similar to the AMM, with the main difference being that the IIM gives a transmission of one up to the critical angle (which once averaged over all possible angles gives an average transmission less than one). The IIM should thus produce lower $R_C$ than the AMM, as well as the DMM. For example, the DMM transmission coefficients are \cite{Reddy2005} $\mathcal{T}_{1 \rightarrow 2}=M_2/(M_1+M_2)$ and $\mathcal{T}_{2 \rightarrow 1}=M_1/(M_1+M_2)$ (using the definition $M$ in \cite{defs}), which once inserted into Eq. (\ref{rc}) gives $R_C = (R_2^{\rm ball}+R_1^{\rm ball})/2$ in the case of linear dispersions (larger than the IIM value of $R_C = (R_2^{\rm ball}-R_1^{\rm ball})/2$). We note that our approach is similar to the radiation limit \cite{Stoner1993}, but does not rely on a linear dispersion assumption, correctly predicts $R_C=0$ for two identical materials, and works in the $\mathcal{T}\rightarrow1$ limit.
   
One approximation of this model is that it assumes both materials have states that overlap in $q_{\perp}$ (as considered in Fig. \ref{fig7}b). At a given $\epsilon$, the phonon states in both materials may be in different regions of the Brillouin zone, which could result in a transmission of zero if they do not overlap in $q_{\perp}$. However the IIM would predict a finite transmission since it only considers the number of conducting channels on both sides. Lastly, we note that the IIM does not discriminate among phonon polarizations; it focuses solely on the availability of final states for phonons.

\section{DFT modeling details (bulk phonons)}
\label{app:dft_details}
The phonon dispersions and scattering rates for Si and Ge were computed using DFT. Structural relaxation was performed with the Quantum Espresso (QE) code \cite{QE1,QE2} using norm-conserving Perdew-Burke-Ernzerhof \cite{PBE96} exchange-correlation pseudopotentials. The converged plane-wave energy cutoffs are 60 Ry and 70 Ry for Si and Ge, respectively. With a Brillouin zone sampling of 11$\times$11$\times$11, the converged lattice parameters are 5.574{\AA} and 5.772{\AA}, respectively, which are 2.6\% and 2.0\% larger than experiment. The second-order interatomic force constants were calculated as follows. With Si, density functional perturbation theory was employed, as implemented in QE, using a 4$\times$4$\times$4 $q$-grid. With Ge, the finite-displacement method was adopted, using Phonopy \cite{Phonopy}, with force calculations carried out on a 5$\times$5$\times$5 supercell and with a $k$-grid sampling the $\Gamma$-point. For anharmonic third-order force constant calculations, performed with ShengBTE \cite{ShengBTE}, a 4$\times$4$\times$4 supercell was employed in both materials. Interactions of up to fourth (Si) and seventh (Ge) nearest neighbors were considered. The phonon dispersions and 3-phonon anharmonic scattering rates, used to calculate $M_{\rm Si,Ge}(\epsilon)$, $\lambda_{\rm Si,Ge}(\epsilon)$ and $g_{\rm Si,Ge}(\epsilon)$, were computed on a $q$-grid of 60$\times$60$\times$60.


\end{document}